\documentclass[twocolumn,showpacs,preprintnumbers,amsmath,amssymb,pre]{revtex4}

\usepackage{graphicx}
\usepackage{dcolumn}
\usepackage{bm}
\usepackage[]{epsfig}
\renewcommand{\phi}{\varphi}

\begin{document}
\title{Ultra-long range correlations of the dynamics of jammed soft matter}
\author{S. Maccarrone$^1$, G. Brambilla$^1$, O. Pravaz$^1$, A. Duri$^1$, M. Ciccotti$^1$, J.-M. Fromental$^1$, E. Pashkovski$^2$, A. Lips$^3$, D. Sessoms$^4$, V. Trappe$^4$, L. Cipelletti$^1$}

\affiliation{$^1$Laboratoire des Collo{\"\i}des, Verres et
Nanomat{\'e}riaux, UMR 5587, Universit{\'e} Montpellier II and CNRS,
34095 Montpellier, France\\
$^2$Unilever Res Labs, Trumbull, CT 06611 USA \\
$^3$Unilever Discover, Port Sunlight, UK \\
$^4$Physics Department, Universit\'e de Fribourg, 1700 Fribourg,
Switzerland\\}

\email{lucacip@lcvn.univ-montp2.fr}
\date{\today}

\begin{abstract}
We use Photon Correlation Imaging, a recently introduced
space-resolved dynamic light scattering method, to investigate the
spatial correlation of the dynamics of a variety of jammed and
glassy soft materials. Strikingly, we find that in deeply jammed
soft materials spatial correlations of the dynamics are quite
generally ultra-long ranged, extending up to the system size, orders
of magnitude larger than any relevant structural length scale, such
as the particle size, or the mesh size for colloidal gel systems.
This has to be contrasted with the case of molecular, colloidal and
granular ``supercooled'' fluids, where spatial correlations of the
dynamics extend over a few particles at most. Our findings suggest
that ultra long range spatial correlations in the dynamics of a
system are directly related to the origin of elasticity.  While
solid-like systems with entropic elasticity exhibit very moderate
correlations, systems with enthalpic elasticity exhibit ultra-long
range correlations due to the effective transmission of strains
throughout the contact network.
\end{abstract}

\maketitle


\section{Introduction}

Soft materials such as colloidal suspensions, emulsions and
surfactant phases typically exhibit increasingly slow relaxation
dynamics as a result of particle crowding or because of strong
interactions, either attractive or
repulsive~\cite{CipellettiJPCM2005}. Work in the past years has
shown that a feature shared by most of these systems is the
heterogenous character of their slow dynamics, which result from
rearrangements that are localized in space and intermittent in
time~\cite{KegelScience2000,WeeksScience2000,LucaJPCM2003,CipellettiJPCM2005,SarciaPRE2005,BerthierScience2005,KilfoilXXXPRL,SalomonXXXPRE,BallestaNatPhys2008,DuriEPL2006},
in analogy with molecular glass
formers~\cite{GlotzerNonCrystSolids2000,EdigerAnnuRevPhysChem2000}
and driven athermal grains and
foams~\cite{DauchotPRL2005_2,DauchotPRL2009XXXcheck,KeysNaturePhysics2007,DurianPREGrains,MayerPRL2004}.

For colloidal hard spheres, probably the most studied soft matter
model system exhibiting a glass transition~\cite{PuseyNature1986},
the size $\xi$ of dynamical clusters undergoing correlated
rearrangements has been shown to grow with volume fraction on
approaching the glass
transition~\cite{WeeksScience2000,WeeksJPCM2007,BerthierScience2005,BrambillaPRL2009}.
For supercooled samples whose dynamics are stationary, the growth is
however modest, the largest reported values of $\xi$ being of the
order of a few particle
sizes~\cite{WeeksScience2000,WeeksJPCM2007,BerthierScience2005,BrambillaPRL2009}.
Similar results have been reported for weakly attractive
systems~\cite{KilfoilXXXPRL,SalomonXXXPRE}. This agrees with
numerical and experimental findings for molecular glass
formers~\cite{GlotzerNonCrystSolids2000,EdigerAnnuRevPhysChem2000,DalleFerrierPRE2007,karmakar09}
and
grains~\cite{DauchotPRL2005_2,DauchotPRL2009XXXcheck,KeysNaturePhysics2007,DurianPREGrains}.
Although the debate is still very active on whether or not the glass
and the jamming transitions coincide in thermal hard
spheres~\cite{LiuKamienPRL,BerthierEPL2009,BerthierPRE2009,xu09,zhang09,jacquin10},
and thus on how far the regime where $\xi$ grows may extend, it is
unlikely that significantly larger correlation lengths may be
measured in the supercooled regime.

Recent experiments using time-resolved light scattering methods
suggest that this scenario could be very different for a wide range
of soft materials quenched in a nearly-arrested, out-of-equilibrium
state, which we shall refer to (somehow loosely) as jammed soft
systems. In these experiments, very large temporal fluctuations of
the intensity autocorrelation function were
observed~\cite{BallestaNatPhys2008,DuriEPL2006,BissigPhysChemComm2003},
suggesting that the size of regions undergoing correlated
rearrangements may be a sizeable fraction of the scattering volume,
i.e. that they may extend over macroscopic distances. Indeed, direct
measurements of $\xi$ in a strongly attractive colloidal
gel~\cite{DuriPRL2009} and in concentrated soft
particles~\cite{SessomsPhilTransA2009} have shown that in these systems the correlation length of the dynamics is limited essentially only by the system
size.

In this paper, we present data on a variety of soft jammed systems,
showing that extremely long ranged correlations of the dynamics in
jammed systems are the rule rather than the exception. The systems
investigated include hard and soft spheres, colloidal gels made of
attractive particles, biomimetic protein films, a concentrated
surfactant solution (``onion'' phase) and Laponite suspensions: with the exception of
the supercooled hard spheres, $\xi$ always exceeds 1 millimeter,
much larger than any structural length scale. We discuss the role of
both the strength and the microscopic origin of the elasticity in
shaping spatial correlations of the dynamics and compare our results
to the behavior of jammed materials under
shear~\cite{LechenaultEPL2008,OlssonPRL2007,HeussingerPRL2009,Heussingercondmat2010}.


\section{Materials and methods}

\subsection{Time resolved correlation and photon correlation imaging }
\label{sec:PCI}

Although dynamic light scattering (DLS) is now a popular and well
established technique to probe the dynamics of soft
materials~\cite{Berne1976}, its usefulness to measure dynamical
heterogeneity has been limited until recently by the spatial and
temporal averages usually involved in DLS measurements. In this
section, we briefly recall the main features of a series of DLS
time- and space-resolved methods that we have introduced in the past
years. More details can be found in
Refs.~\cite{LucaJPCM2003,BissigPhysChemComm2003,DuriFNL2005,DuriPRE2005,DuriPRL2009}.

\textit{Time resolved correlation (TRC)}. Valuable information on
spatial correlations of the dynamics can be obtained in
time-resolved measurements, even if they lack spatial resolution.
Intuitively, this is due to the fact that temporal fluctuations of
the dynamics are enhanced if the probed sample volume contains a
limited number of statistically independent regions, as is the case
when the dynamics are correlated over large
distances~\cite{DurianPREGrains}. More formally, a dynamical
susceptibility $\chi_4$ can be introduced~\cite{Parisichi4}, which
quantifies temporal fluctuations of the dynamics, and which is
proportional to the volume integral of the spatial correlation of
the dynamics, $G_4$~\cite{Parisichi4,LacevicPREXXXcheck}.
In our DLS experiments, we achieve temporal resolution by using a
CCD camera as a detector and by averaging the intensity correlation
function over pixels rather than over time. The CCD is placed in the
far field, so that each pixel is illuminated by light issued from
the whole scattering volume at a well defined scattering angle
$\theta$. The dynamics are probed on a length scale $\ell \sim 1/q$,
where $q = 4\pi n \lambda^{-1} \sin(\theta/2)$ is the scattering
vector and $n$ and $\lambda$ are the solvent refractive index and
the in-vacuo wave length of the laser that illuminates the sample,
respectively. We measure a time-resolved intensity correlation
function (proportional to the square of the dynamic structure
factor), defined as
\begin{equation}
c_I(t,\tau) = \frac{\left < I_p(t)I_p(t+\tau) \right >_p}{\left <
I_p(t) \right >_p \left < I_p(t+\tau) \right >_p} -1 \,,
\label{eq:cI}
\end{equation}
which we refer to as the degree of correlation. In
Eq.~(\ref{eq:cI}), $I_p$ indicates the scattered intensity measured
by the $p$-th pixel and the averages are taken over the whole CCD
detector. The average dynamics and dynamical heterogeneity are
quantified by
\begin{eqnarray}
g_2(\tau)-1 = <c_I(t,\tau)>_t \\
\chi(\tau) = var[c_I(t,\tau)] = <c_I(t,\tau)^2>_t -
<c_I(t,\tau)>^2_t\,, \label{eq:g2andchi}
\end{eqnarray}
where averages are taken over time and where $g_2-1$ is the usual
intensity correlation function measured in DLS, while $\chi$ is the
equivalent for DLS of the dynamical susceptibility $\chi_4$ used in
numerical works. Note however that, contrary to $\chi_4$, $\chi$ is
not normalized with respect to the number of particles in the
scattering volume, a quantity not always easily accessible
experimentally.

\begin{figure}
  \psfig{file=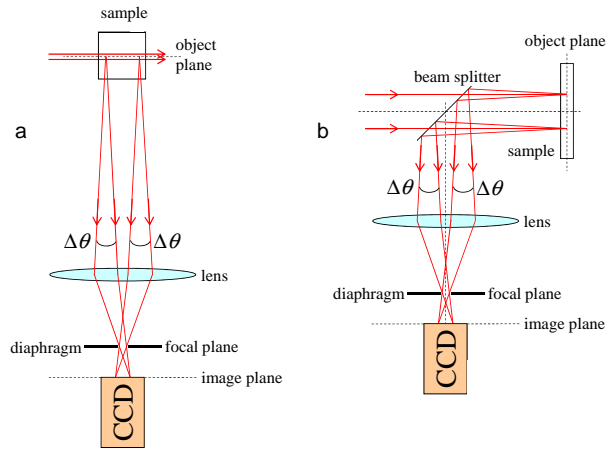,width=8.cm}
  \caption{Scheme of a PCI setup in two different configurations: a) $90^{\circ}$ scattering angle; b) backscattering. From Ref.~\cite{DuriPRL2009} with permission.}
  \label{fig:setup}
\end{figure}

\textit{Photon correlation imaging (PCI).} Spatial resolution may be
achieved by modifying the collection optics so as to form an image
of the scattering volume onto the CCD detector. This method, which
we have termed photon correlation imaging (PCI), has been described
for a low-$q$ setup in Ref.~\cite{DuriPRL2009}. In
Fig.~\ref{fig:setup} we show its implementation for $\theta = 90$
(a) and 180 (b) degrees, as in the experiments reported here. In
both cases, an image of the illuminated sample is formed onto the
CCD with magnification $M \sim 1$, using light scattered within a
small solid angle centered around a well defined scattering angle.
The resulting image has a speckled appearance similar to that of
conventional far-field scattering experiments. In practice, one
adjusts the diaphragm aperture so as to roughly match the speckle
size to the pixel size, which is typically of the order of
$10~\mu$m. Under these conditions, the speckle size is larger than
the typical size of the scatterers ($\lesssim 1~\mu$m), which thus
can not be resolved individually. Information on the local dynamics,
however, can still be obtained, because the relative motion of the
scatterers results in fluctuations of the intensity of each speckle,
as in regular DLS. In contrast to conventional DLS, here each
speckle is illuminated by light issued from a well defined, small
region of the sample, so that spatially-resolved dynamics can be
measured.

In practice, we divide the CCD images in regions of interest (ROIs),
each corresponding to a small volume in the sample, and apply the
TRC method separately to each ROI. We compute a space- and
time-resolved degree of correlation, defined as
\begin{equation}
c_I(t,\tau,\mathbf{r}) = \frac{\left < I_p(t)I_p(t+\tau) \right
>_\mathbf{r}}{\left < I_p(t) \right >_{\mathbf{r}} \left < I_p(t+\tau) \right >_{\mathbf{r}}} -1
\,, \label{eq:cIspaceresolved}
\end{equation}
where the average is now over all pixels within a ROI corresponding
to a small volume centered around $\mathbf{r}$. The ROIs must
contain at least about 100 speckles for the statistical noise on the
local degree of correlation to be acceptable: this limits the
resolution of the method, which should be regarded as a
coarse-grained technique. The main advantage of PCI is to decouple
the size of the field of view (dictated by the magnification $M$)
from the length scale $\ell$ over which the dynamics are probed
(dictated by $q$ and thus the scattering angle). Thus, very
restrained motion can be measured with a large field of view. This
is not the case for conventional imaging methods, where $\ell$ is a
fixed fraction of the field of view.

The spatial correlation of the dynamics may be quantified by
comparing the temporal evolution of the local degree of correlation
calculated, for a given lag, for different locations. Traces
corresponding to pairs of regions with correlated dynamics will
exhibit similar fluctuations, while uncorrelated ROIs yield
independent fluctuations of $c_I$. An example that will be discussed
towards the end of this paper is given in Fig.~\ref{fig:laponite}.
We define
\begin{equation}
\widetilde{G_4}(\tau,\Delta \mathbf{r}) = \left < \frac{\left<\delta
c_I(t,\tau,\mathbf{r})\delta
c_I(t,\tau,\mathbf{r}+\Delta\mathbf{r})\right
>_t}{\sqrt{var\left[c_I(t,\tau,\mathbf{r})\right]var\left[c_I(t,\tau,\mathbf{r}+\Delta\mathbf{r})\right]}}
\right >_\mathbf{r} \,,
 \label{eq:G4tilde}
\end{equation}
where $\delta c_I = c_I - <c_I>_t$ are the temporal fluctuations of
the local dynamics. This is the analogous, albeit at a coarse
grained level, of the spatial correlation of the dynamics calculated
in numerical and experimental work where particle trajectories are
accessible~\cite{LacevicPREXXXcheck,DauchotPRL2005_2,LechenaultEPL2008,WeeksJPCM2007}.
In most cases, the dynamics are isotropic and we average
$\widetilde{G_4}$ over all orientations of $\Delta \mathbf{r}$.

An important point concerns the normalization of
Eq.~(\ref{eq:G4tilde}): we recall that $c_I$ contains a noise
contribution due to the statistical noise associated with the finite
number of pixels processed for each ROI~\cite{DuriPRE2005}. In PCI
experiments, this contribution may be quite large, since one
typically tries to reduce the size of the ROIs as much as possible
in order to achieve a better spatial resolution. Distinct ROIs have
uncorrelated noise; as a result, the numerator of the r.h.s. of
Eq.~(\ref{eq:G4tilde}) is noise-free (except for $\Delta r=0$),
while the noise contributes to the denominator by increasing its
value. Thus, at spatial lags $>0$ $\widetilde{G_4}$ is depressed
because of the noise contribution, an effect that depends on the
size of the ROIs used for the analysis. For systems that exhibit a
finite spatial correlation of the dynamics at $\Delta r > 0$, we
introduce a normalized spatial correlation of the dynamics defined
as
\begin{equation}
G_4(\tau,\Delta \mathbf{r}) = b(\tau)\widetilde{G_4}(\tau,\Delta
\mathbf{r})\,,
 \label{eq:G4}
\end{equation}
where $b$ is a time-lag dependent coefficient chosen so that $G_4
\rightarrow 1$ for $\Delta \mathbf{r} \rightarrow 0$.

\subsection{Experimental systems}

We report below PCI measurements of heterogeneous slow dynamics on a
variety of systems: ``artificial skin'' (Vitro-corneum$^{\circledR}$
by IMS Inc.), concentrated colloidal hard spheres, an ``onion'' gel,
and a Laponite suspension. For the sake of comparison, we recall
also results obtained from concentrated soft spheres and a colloidal
gel, taken from Refs.~\cite{SessomsPhilTransA2009} and
~\cite{DuriPRL2009}, respectively. For all systems, data are taken
under single scattering conditions. The experiments on the
``artificial skin'' will be discussed somehow more in depth, because
\textit{i)} its behavior is representative of the main features
observed for the other systems; \textit{ii)} unlike most colloidal
systems where the control parameters for the slow dynamics are
volume fraction and interparticle potential, in ``artificial skin''
the slow dynamics is controlled by the relative humidity (RH). In this respect, ``artificial skin'' is representative of a class of materials of biological relevance whose slow
dynamics are still poorly characterized and whose visco-elastic
properties change dramatically with changes in hydration,
as in stratum corneum, the uppermost part of the epidermis of mammalian
skin~\cite{skin1,skin2,skin3}. This behavior is due to the fact that, for these materials, water acts as a plasticizer, lowering their glass transition temperature below room temperature at high RH.

The ``artificial skin'' is a protein-based thin film ($20-25~\mu$m)
that mimics the properties of human stratum corneum. We mount a piece of the film on a circular frame of
10 mm of diameter to keep it flat and place it in a sealed
custom-made cell that allows the relative humidity (RH) to be
controlled during the PCI measurements. The desired value of the RH
is imposed by placing a saturated solution of a suitable salt in a
reservoir contained in the cell. We present here results for two
saturated salt solutions (prepared by adding 20\% more in weight
fraction than the solubility limit at $20^{\circ}$C), yielding a
humid (RH = 62\%, using KI) and a dry (RH = 12\%, using LiCl)
atmosphere, respectively. The measurements are performed in the
backscattering geometry depicted schematically in Fig. 1b (where the
RH-controlling cell is not shown for simplicity), with $\theta =
180$ degrees, corresponding to a typical probed length scale $\ell
\sim 30$ nm. The imaged portion of the sample has a size of $4.288
\times 4.288~\mathrm{mm}^{2}$ ($3.752 \times 3.752~\mathrm{mm}^{2}$)
for the dry (humid) sample; in both cases, square ROIs of size $268
\times 268 ~\mu\mathrm{m}^{2}$ were used for the PCI analysis. More
details on the lags used and the duration of the experiment are
given in the discussion below.

The colloidal hard sphere samples are suspensions of poly-(methyl
methacrylate) spheres of radius $\sim 100$~\cite{brambilla10} nm in
an index matching mixture of cis/trans-decalin and tetralin. Their
average dynamics have been reported in
Refs.~\cite{BrambillaPRL2009,ElMasriJSTAT2009}. For the
space-resolved measurements discussed here, we use the PCI geometry
shown in Fig.~\ref{fig:setup}a, where $q = 25~\mu\mathrm{m}^{-1}$,
and where the size of the field of view and ROIs are $2.5 \times
0.6~\mathrm{mm}^{2}$ and $55 \times 55 ~\mu\mathrm{m}^{2}$,
respectively. We present data for two volume fractions, $\varphi =
0.5468$ (delay time $\tau = 20$ ms, relaxation time of the dynamic
structure factor $\tau_{\alpha} = 0.147$ s) and $\varphi = 0.5957$
($\tau = 14$ s, $\tau_{\alpha} = 2350$ s). All data are taken in a
regime were the dynamics are stationary.

The onion gels are concentrated surfactant solutions forming a dense
packing of polydisperse, deformable spheres of average size of the
order of a few $\mu\mathrm{m}$~\cite{RamosEPL2004}. Each sphere is
constituted by a stacking of surfactant bilayers that roll up to
form a multilamellar vesicle, or onion. The average dynamics and the
rheological properties of the onions have been described in
Refs.~\cite{RamosPRL2001,RamosPRL2005}; an optical microscopy
investigation~\cite{MazoyerPRL2006,MazoyerPRE2009} suggests that
their dynamics is correlated at least up to length scales comparable
to the field of view accessible in those experiments, about 1 mm. In
the PCI measurements reported here, we use the $90$ degree
scattering angle geometry of Fig.~\ref{fig:setup}a, with a field of
view of $1.9 \times 0.39~\mathrm{mm}^{2}$ and ROIs of size $130
\times 130 ~\mu\mathrm{m}^{2}$. The delay time is $\tau = 1000$ s,
about 20 times smaller than the relaxation time of $g_2-1$. The
dynamics slow down with time: we analyze data for $20000 < t< 60000$
s ($t=0$ being the time at which the surfactant solution is quenched
in the gel phase by a temperature jump~\cite{RamosPRL2001}).

Laponite RD (Rockwood, US) is a synthetic clay consisting of discoid
charged particles. By dispersing 3.5\%wt of Laponite powder into
pure water, we obtain a colloidal glassy suspension that keeps aging
for several days~\cite{MourchidLangmuir1995,KnaebelEPJL2000}. To properly weight the Laponite content, we dry
the powder for 12 hours in an oven at 130 C$^{\circ}$. In order to
prevent chemical corrosion, the Laponite powder is then dispersed in
a pH 10 solution constituted of pure Millipore water and a proper
amount of sodium hydroxide. The white dispersion is held in a
magnetic stirrer for 25 minutes until it becomes almost transparent,
and it is then injected into the scattering cell through a 1
$\mu\mathrm{m}$ filter in order to eliminate particle clusters. The
sample is prepared in an inert argon atmosphere in order to prevent
contact with $\mathrm{CO}_2$. The scattering cell is then sealed in
order to preserve this pure condition. The final step before
measurements consists of a 3 minute centrifugation at 3000 rpm, to
eliminate small gas bubbles that may have formed during the process.
This sets the age zero of the sample $(t=0)$ with an incertitude of
a few minutes. The data shown here refer to $2.0 \times
10^5~\mathrm{sec}~< t < 2.5 \times 10^5~\mathrm{sec}$; they were
taken using a special cell as described at the end of the next
section.

\section{Results}

\begin{figure}
  \centering
  \psfig{file=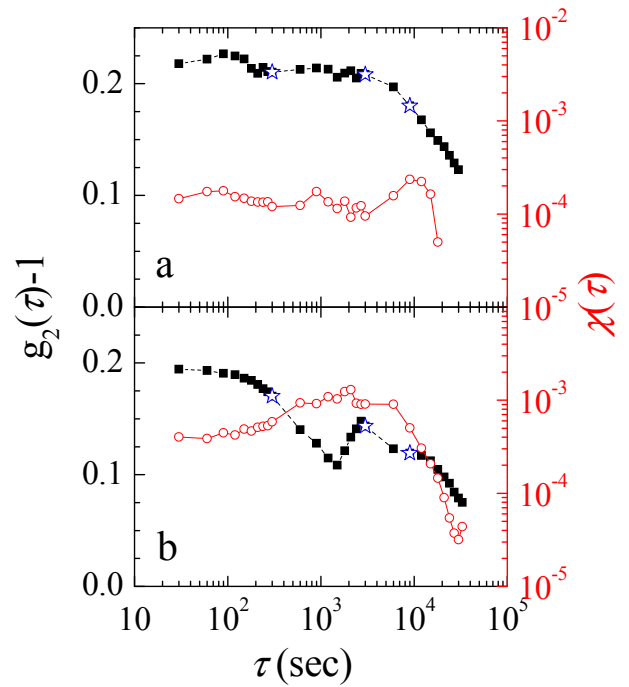,width=8.cm}
  \caption{Intensity time autocorrelation function, $g_2-1$ (left axis and solid black squares),
  and dynamical susceptibility, $\chi$ (right axis and open circles), for the
  ``artificial skin'' sample, as measured in the backscattering configuration schematized
  in Fig.~\ref{fig:setup}b. a: relative humidity = 62\%; b: relative humidity =  12\%.
  Note the enhanced dynamical heterogeneity in dry atmosphere, as shown by the higher values of $\chi$.
  The three lags indicated by open stars are those for which the spatial correlation of the dynamics
  is shown in Fig.\ref{fig:ASG4}.}
  \label{fig:ASg2}
\end{figure}

As a representative example of dynamical heterogeneity in jammed
systems, we first describe in some detail the behavior of
``artificial skin'' films. We show in Fig.~\ref{fig:ASg2} both the
average dynamics and the dynamical susceptibility for a film in
humid (RH = 62\%, a) and dry (RH = 12\%, b) atmosphere. The data
shown here are obtained by processing the full images and by
averaging over several tens of thousands of sec (36000 sec for the
sample at RH = 12\% and 22500 sec for RH = 62\%). The temporal
intensity correlation functions (black squares, left axis) exhibit a
decay on a very long time scale, of the order of $10^4$ sec. For the
dry sample, $g_2-1$ appears to be more noisy than for the humid one,
with an additional decay at earlier times, $\tau~\approx 400$ sec.
This ``noisiness'' is due to enhanced dynamical heterogeneity, as
compared to the humid sample: as discussed below, the dynamics of
the dry sample are intermittent in time and correlated in space over
very large distances: as a result, the temporal and spatial averages
performed in calculating $g_2-1$ are poorer than for the humid
sample. In both panels, we show also the dynamical susceptibility
$\chi(\tau)$ (open circles and right axis). For the humid sample,
$\chi$ depends weakly on $\tau$, barely exhibiting a peak on the
time scale of the decay of $g_2-1$. A peak in $\chi$ is a
distinctive feature of dynamical
heterogeneity~\cite{LacevicPREXXXcheck,karmakar09,DauchotPRL2005_2,DuriPRE2005},
while the noise contribution to the dynamical susceptibility is
essentially flat for time delays smaller than the decay time of the
intensity correlation function~\cite{DuriFNL2005,DuriPRE2005}. Thus,
the data in Fig.~\ref{fig:ASg2}a suggest that the dynamics for the
humid sample are mildly heterogeneous and that the noise
contribution dominates the measured $\chi$, with the exception of
the small peak around $\tau = 10^4 $ sec. By contrast, the dynamical
susceptibility of the dry sample presents a well-developed peak,
about a factor of three higher than the short-$\tau$ value of the
dynamical susceptibility and more than a decade higher than $\chi$
at large delay times. This suggests that the dynamics of the dry
sample are highly heterogeneous.

\begin{figure}
  \psfig{file=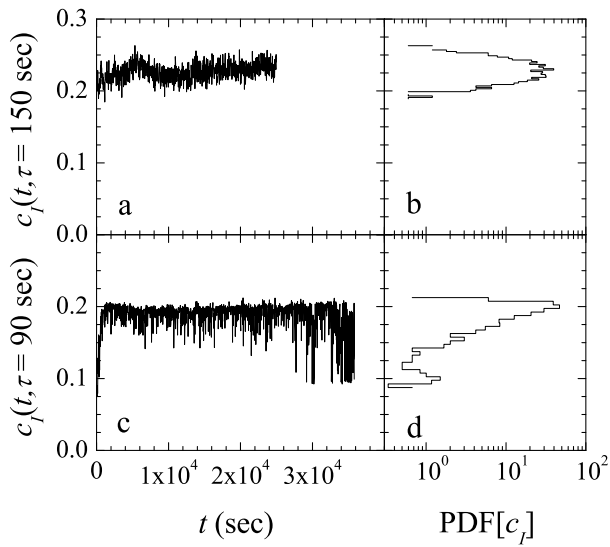,width=8.cm}
  \caption{For a representative ROI, degree of correlation $c_I$ at a fixed delay $\tau$
  as a function of time and its PDF for the ``artificial skin'' in humid (
  RH = 62\%, $\tau = 150$ sec, panels a and b) and dry (RH = 12\%, $\tau = 90$ sec,
  panels c and d) atmosphere. Note that the dry sample exhibits large temporal fluctuations.}
  \label{fig:AScI}
\end{figure}

In order to gain further insight on the nature of the dynamical
heterogeneity of ``artificial skin'', we plot in Fig.~\ref{fig:AScI}
the temporal evolution of the degree of correlation, $c_I$,
calculated for a time delay much smaller than the relaxation time of
$g_2$ ($\tau = 150$ and $90$ sec for the humid and dry sample,
respectively). We focus on small delays because this allows us to
capture better the impact of individual rearrangement events, since
at larger $\tau$ several events may occur in between two images. The
signal shown here have been calculated for an individual ROI, but
are representative of the general behavior of all ROIs. For the
humid sample, $c_I$ exhibits modest fluctuations symmetrically
distributed around its mean value (panel a). Moreover, the
probability distribution function (PDF) of $c_I$, shown in panel b,
is close to Gaussian. We have shown in previous
work~\cite{DuriFNL2005,DuriPRE2005,BissigPhysChemComm2003} that this
is typical of a signal dominated by the measurement noise, which is
particularly high when the degree of correlation is calculated over
the limited number of pixels of a small portion of the image, as
required by a space-resolved analysis. Panels c and d show $c_I$ and
its PDF, respectively, for the dry sample. In contrast to the humid
film, sudden drops of the degree of correlation can be easily
observed, corresponding to sudden rearrangement events in the
sample. As a result the PDF of the degree of correlation is skewed,
with a tail associated with the intermittent drops of $c_I$. In
other systems, the shape of the PDF has been shown to be well fitted
by a Gumbel distribution~\cite{DuriPRE2005}, often observed in the
statistics of rare events in systems with extended spatial and
temporal correlations~\cite{GumbellstuffXXX}. Here, the dynamics
were too slow to allow us to accumulate enough statistics to make
any quantitative statement on the shape of the PDF. We note however
that the analysis of the temporal evolution of $c_I$ and its PDF
confirms the differences between the two samples discussed in
relation to Fig.~\ref{fig:ASg2}: the dynamics of the dry sample are
much more heterogeneous than those of the humid one.

\begin{figure}
  \psfig{file=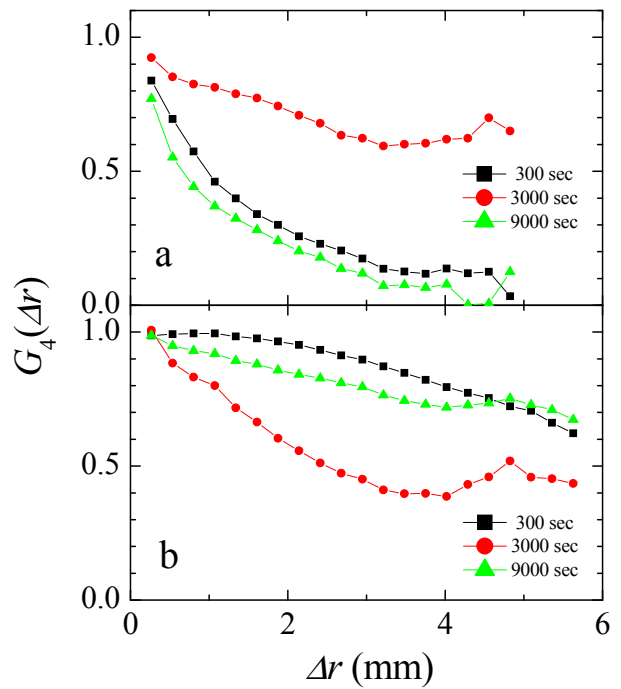,width=8.cm}
  \caption{Spatial correlation of the dynamics, $G_4(\Delta r)$, for the ``artificial skin''
  samples in humid (RH = 62\%, a) and dry (RH = 12\%, b) atmosphere. The curves are labeled by the
  temporal lag at which the dynamics were probed.}
  \label{fig:ASG4}
\end{figure}

Since the dynamics measured on the full field of view
(Fig.~\ref{fig:ASg2}) exhibit significant fluctuations, we expect
them to be spatially correlated over macroscopic distances. The
spatial correlation of the dynamics calculated according to
Eqs.~(\ref{eq:G4tilde}) and (\ref{eq:G4}) is shown in
Fig.~\ref{fig:ASG4}, for the two samples, and for the three time
lags indicated by open stars in Fig.~\ref{fig:ASg2}. Quite
generally, $G_4$ is significantly larger than zero up to distances
of the order of 1 mm (for the humid sample) or even several mm (for
the dry sample). This is remarkable, since all relevant structural
length scales are much smaller. Furthermore, the range of the
spatial correlations of the dynamics is much longer for the dry
sample than for the humid one; together with the enhanced
intermittency under dry conditions seen in Fig.~\ref{fig:AScI}, this
explains why the dynamical susceptibility of the dry sample is
higher than that of the wet one, as shown in Fig.~\ref{fig:ASg2}.
Interestingly, the time lag dependence of $G_4$ is different for the
two values of RH. At high RH, the range of $G_4$ is maximum at
intermediate time delays and decreases at small and large $\tau$.
This is similar to what reported in numerical work on supercooled
molecular glass formers~\cite{LacevicPREXXXcheck} and experiments on
grains~\cite{DauchotPRL2009XXXcheck}, suggesting that dynamical
heterogeneity progressively build up with time, starting from
individual events that are relatively limited in size. Eventually,
at very large time lags, one expects $G_4$ to decay on short length
scales, since many uncorrelated events will have occurred everywhere
in the sample. The shorter range of $G_4$ at $\tau = 9000$ sec
observed in Fig.~\ref{fig:ASG4}a probably corresponds to the onset
of this large $\tau$ regime. For the dry sample in panel b, $G_4$ is
highly correlated over a very long range for essentially all probed
lags, indicating that individual events extend over several
millimeters. We don't have currently an explanation for the somehow
faster decay of $G_4$ at intermediate lags; the long range of $G_4$
at the largest probed lag ($\tau = 9000$ sec) indicates that on this
time scale the regime where many uncorrelated events have occurred
throughout the sample is not yet attained. Longer lags could not be
reliably analyzed due to the limited duration of the experiment.

Summarizing, the dynamics of the ``artificial skin'' are markedly
heterogeneous under dry conditions, associated with intermittent
rearrangement events. These events imply the motion of the
scatterers over just a few nanometers (we recall that the full decay
of $g_2-1$ corresponds to displacements of the order of 30 nm);
however, whenever they occur, they impact essentially the whole
sample, since the spatial correlation of the dynamics extends over
several mm. For the humid sample, dynamical heterogeneities are less
pronounced; the range of spatial correlations, although reduced in
comparison to that of the dry sample, is still on the order of 1-2
mm, much larger than any structural length scale in the sample. The
change in dynamical behavior under different RH conditions observed
here for the ``artificial skin'' is related to the loss of rigidity
upon hydration and is reminiscent of the sensitivity to humidity of
biological tissues. Indeed, humidity plays a key role in controlling
biomechanical properties of stratum corneum~\cite{Eugene1}, whose
elastic modulus, e.g., increases by several orders of magnitude with
decreasing RH~\cite{skin2,skin3}.

\begin{figure}
  \centering
  \psfig{file=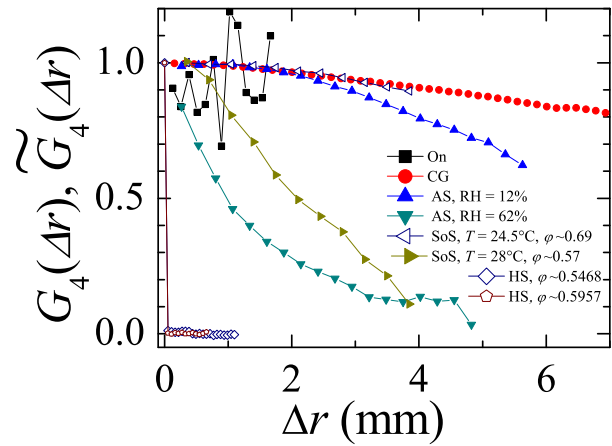,width=8.cm}
  \caption{Spatial correlation
of the dynamics for all systems that we have studied except for
Laponite, as indicated by the label (On: onion gel; CG: colloidal
gel, from Ref.~\cite{DuriPRL2009}; AS: ``artificial skin'',
$\tau=300$ sec; SoS: soft spheres,
from~\cite{SessomsPhilTransA2009}; HS: hard spheres). Data for
Laponite are shown in Fig.~\ref{fig:laponite} below. All spatial
correlation functions have been normalized according to
Eq.~(\ref{eq:G4}), except for the hard spheres, for which
$\widetilde{G_4}$ is shown (see Eq.~(\ref{eq:G4tilde})), rather than
$G_4$.}
  \label{fig:G4all}
\end{figure}

One may wonder whether similar dynamical features are observed also
in other glassy or jammed soft matter systems that, similarly to the
``artificial skin'', are characterized by ultraslow dynamics and a
viscoelastic behavior where the solid-like response dominates. To
address this question, we plot in Fig.~\ref{fig:G4all} the spatial
correlation function of the dynamics of a series of systems:
concentrated hard spheres at the onset of the supercooled regime and
in the deep supercooled regime, soft spheres below and above random
close packing (data taken form Ref.~\cite{SessomsPhilTransA2009}), an
onion gel, a colloidal gel (data taken from
Ref.~\cite{DuriPRL2009}), and the dry ``artificial skin'' film
already shown in Fig.~\ref{fig:ASg2}. With the exception of the soft
spheres and the colloidal gel, for which data were taken in a
low-angle configuration ($q \sim 1~\mu\mathrm{m}^{-1}$), all data
are collected either at $\theta = 90$ or $180$ degrees. For
all samples, we calculate the spatial correlation of the dynamics at
a time lag $\tau$ much shorter than the relaxation time of the
corresponding $g_2-1$, in order to capture as much as possible the
characteristics of single events. Quite generally, very long range
correlations of the dynamics are observed for all samples, up to
several mm, once again much larger than any structural length scale
and comparable to the system size. In particular, data for the
onions, albeit somehow noisy due to the relatively small number of
available pixels and the limited duration of the experiment (about
twice the relaxation time of $g_2-1$), confirm and extend the long-range
correlation of the dynamics observed, for a smaller field of view,
by microscopy~\cite{MazoyerPRL2006,MazoyerPRE2009}. Even for the
``artificial skin'' under humid conditions and the less compressed
soft sphere system ($T = 28^{\circ}$C), where $G_4$ decreases more
rapidly than for the other systems, spatial correlations still extend
over macroscopic length scales. The only exceptions are the hard
spheres samples. For these systems, $\widetilde{G_4}$ drops to zero
as soon as $\Delta r
> 0$; although we show here data at $\tau$ much smaller than the relaxation time of $g_2-1$,
we point out that a similar behavior is found at all $\tau$. Since
the normalization method described in Sec.~\ref{sec:PCI} can not be
applied, we represent $\widetilde{G_4}$ rather than $G_4$ as for the
other samples. For the less concentrated hard sphere suspension,
such a sharp drop of $\widetilde{G_4}$ is not surprising, since no
spatial correlation of the dynamics are to be expected at the onset
of the supercooled regime. Indeed, at $\phi = 0.5468$ the dynamics
is slower than in the $\phi \rightarrow 0$ limit by a relatively
modest factor of 100, and the shape of the dynamic structure factor
differs only marginally from that in the diluted case~\cite{BrambillaPRL2009}. For the most
concentrated hard sphere sample, by contrast, the system relaxation
is already slowed down by a factor of about $10^7$ compared to the
$\varphi \approx 0$ limit and a fully developed plateau is observed
in the dynamic structure factor, revealing caging and glassy
dynamics~\cite{BrambillaPRL2009,ElMasriJSTAT2009}. It has been
proposed that such a dramatic slow down of the dynamics is
associated with the growth of spatial correlations of the dynamics.
Previous experiments, both by confocal
microscopy~\cite{WeeksJPCM2007} and dynamic light
scattering~\cite{BrambillaPRL2009,BerthierScience2005}, indicate
that the range, $\xi$, of such correlations does not exceed a few
particles sizes, i.e. a few $\mu$m in our case. Therefore, the
absence of any measurable correlation in our PCI experiment, where
the smallest accessible spatial lag is $55~\mu\mathrm{m}$, is fully
consistent with previous results. On the one hand, this negative
result illustrates the limitations intrinsic to PCI, a coarse
grained method. On the other hand, it highlights the dramatic
difference between the dynamics of supercooled, stationary hard
spheres and those of the other jammed, out-of-equilibrium samples,
for which macroscopic spatial correlations of the dynamics are
observed.

\begin{figure}
  \centering
  \psfig{file=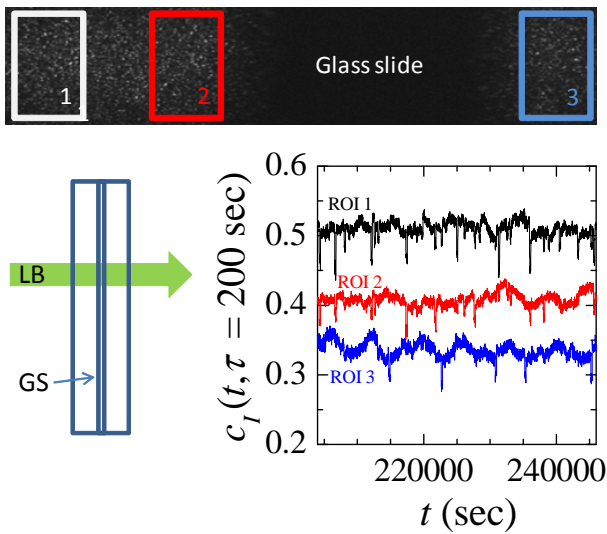,width=8.cm}
  \caption{Bottom left: schematic side view of the cell used for the experiment on Laponite. LB: laser beam, GS: glass slide.
  Top: typical CCD image of the scattering volume. The dark region corresponds to the thickness of the glass slide, view from the side.
  The three ROIs for which the degree of correlation is shown in the bottom right plot are highlighted.
  The size of the imaged region is $2.62 \times 0.52 \mathrm{mm}^2$. Bottom right: time dependence of $c_I$,
  for a delay time $\tau = 200$ sec, for the three ROIs shown above. For the sake of clarity,
  the curves of ROIs 1 and 2 have been offset vertically by 0.2 and 0.1, respectively.
  Note that the signals measured on the same side of the glass slide (ROIs 1 and 2)
  are correlated, while signals from opposite sides are uncorrelated.}
  \label{fig:laponite}
\end{figure}

The results shown here suggest that system-size correlations of the
dynamics are ubiquitous in fully jammed systems, where the solid-like behavior of the sample dominates over its viscous response.
Indeed, in a material that is essentially solid-like, any
deformation due to a local rearrangement will propagate very far
before being substantially damped, leading to extended spatial
correlations of the dynamics. Another (less attractive) explanation
of the dynamical fluctuations reported here could be that they are
due to some kind of artifact. In fact, one should be aware that
measuring fluctuations in a reliable fashion is much more difficult than just probing
the average dynamics, and many spurious effects have to be
controlled and ruled out (for a discussion of some possible sources
of artifacts, such as temperature fluctuations and laser beam
pointing instabilities, see also Ref.~\cite{DuriPRE2005}). In order
to address the role of the elastic propagation of a strain field and
to demonstrate the genuine nature of the spatial correlations
observed here, we discuss briefly an experiment performed on a
Laponite sample using a specially devised cell.

The cell is a cylindrical tube of inner diameter $\approx 10$ mm
(see Fig.~\ref{fig:laponite}), with a microscope glass slide glued
in the interior with epoxy, so as to separate the cell into two
chambers. Because the glass slide is cut unevenly and does not fit
perfectly the cell walls, the two chambers are in contact through
openings of size $\approx 1~$mm, so that the sample contained in the
two chambers is fully equilibrated and experiences exactly the same
conditions. The laser beam impinges perpendicularly to the glass
slide, thereby illuminating both chambers, and the scattering volume
is imaged at an angle $\theta = 90^{\circ}$, as in the geometry shown in
Fig.~\ref{fig:setup}a. We calculate the local $c_I$ for
larger-than-usual ROIs, in order to reduce the noise contribution
(see Fig.~\ref{fig:laponite}). From the plot in
Fig.~\ref{fig:laponite}, it is clear that the time evolution of the
degree of correlation measured on the same side of the glass slide
is highly correlated (ROIs 1 and 2, separated by $\Delta r = 0.57$
mm), since most of the downward spikes corresponding to sudden
rearrangement events are co-occurrent. Indeed, $\widetilde{G_4}$
calculated from the signals for ROIs 1 and 2 is as high as 0.394 (we
recall that $\widetilde{G_4}$ is affected by the noise contribution,
so that an even higher value would be observed in the absence of
noise). By contrast, the events recorded in ROI 3, on the opposite
side of the glass slide, appear to be uncorrelated with respect of
those of ROIs 1 and 2, as confirmed by low values of the correlation
(e.g., $\widetilde{G_4}$ = -0.015 when correlating the signals from
ROIs 2 and 3).

This experiment demonstrates that the downward spikes
observed in $c_I$ can not be due to some mechanical, temperature or
laser beam instability, since they occur independently on either
side of the glass slide. On the other hand, the data of
Fig.~\ref{fig:laponite} support the idea that the propagation of a
strain field in a predominantly elastic medium is responsible for
long range spatial correlations. Indeed, such a strain field can
not propagate through the rigid glass slide, so that the dynamical activity in the two chambers is uncorrelated, while the one in a given chamber is correlated.

\section{Discussion}

\begin{table}
\label{tab:rheoparameters}

\caption{\label{tab:rheoparameters} Rheological parameters of most of the system shown in Figs.~\ref{fig:G4all} and~\ref{fig:laponite}. The sample names are as in the caption of Fig.~\ref{fig:G4all}.}
\begin{ruledtabular}
\begin{tabular}{ccccc}
&$\nu$ (Hz)&$G'(\nu)$ (Pa) & $G'(\nu)/G''(\nu)$ & Ref.\\
\hline
On & 1 & 600 & 15 & \cite{RamosPRL2005} \\
CG & 1 & $\sim 0.9\times 10^{-3}$ & 10 & \cite{NoteRheologyGels} \\
SoS, $\phi = 0.57$&1.6&0.6&0.3&\cite{SessomsPhilTransA2009}\\
SoS, $\phi = 0.69$&1.6&20&8&\cite{SessomsPhilTransA2009}\\
HS, $\phi = 0.5468$&1&40&1.1&\cite{MasonPRLxxx,NoteHS}\\
HS, $\phi = 0.5957$&1&$>80$&$>1.4$&\cite{MasonPRLxxx,NoteHS}\\
Laponite&0.7&$\gtrsim 300$&20&\cite{XXXLaponiteElasticity,NoteLaponite}\\
\end{tabular}
\end{ruledtabular}

\end{table}

To better understand the relationship between viscoelastic
properties and spatial correlations of the dynamics, we examine
published rheological data for most of the systems investigated
here and summarize the relevant rheological parameters in Table~\ref{tab:rheoparameters}, where we choose, somehow arbitrarily, $\nu \sim 1$ Hz as a reference frequency for comparing different systems.
Oscillatory rheology experiments for the onion
gels~\cite{RamosPRL2005} show that both the storage, $G'(\nu)$,
and loss, $G''(\nu)$, moduli are essentially
frequency-independent in the range $4 \times 10^{-3}~\mathrm{Hz} < \nu < 10 ~\mathrm{Hz}$, and that $G' \sim
600~\mathrm{Pa}$ dominates over $G''$ by a factor of $\sim 15$.
Similar results are obtained for the Laponite suspension~\cite{XXXLaponiteElasticity} and for the colloidal gels~\cite{GislerXXXPRL}. Note that the magnitude of the elastic modulus of the colloidal gel ($G'\sim~ 4.4 \times 10^{-4} - 1.5 \times 10^{-3}~\mathrm{Pa}$~\cite{GislerXXXPRL,KrallPRL1998,NoteRheologyGels}) is lower than that of the onion gel and of the Laponite suspension by at least 5 orders of magnitude, implying that the absolute value of $G'$ is not a relevant parameter in determining the range of spatial correlations of the dynamics. A systematic investigation of
the shear moduli of the soft particles is presented in
Ref.~\cite{SessomsPhilTransA2009}. For the most concentrated sample
reported in Fig.~\ref{fig:G4all} above, one finds that $G'$
and $G''$ are nearly frequency independent and that the former is
larger than the latter by almost a decade. By contrast, for the less
concentrated sample both moduli are comparable (see panels a) and
b) of Fig. 3 of Ref.~\cite{SessomsPhilTransA2009}), revealing a
complex viscoelastic behavior where neither the solid-like nor the
fluid-like character prevail. Since spatial correlations of the
dynamics are shorter-ranged in the less concentrated sample, these
observations are consistent with the notion that a fully developed
elastic behavior is a necessary condition for long-range spatial
correlations of the dynamics. The same trend is likely to hold for
the ``artificial skin''. Indeed, we find that the range of the
spatial correlation of the dynamics increases as the RH decreases.
Although no rheology data are available for the ``artificial skin'',
it is worth noting that in stratum corneum~\cite{skin2,skin3}, a
material whose properties the Vitro-corneum$^{\circledR}$  film is
designed to mimic, the elastic modulus grows by almost 3 orders of magnitude when the RH decreases from 100\% to 30\%.

Collectively, these results indicate that systems with long-ranged
spatial correlations of the dynamics have a solid-like behavior. A
close inspection of rheology data for hard sphere suspensions, for
which we recall that $\xi$ is limited to  a few particle sizes at
most, show that the reverse is not true. Mason and Weitz have
measured the shear moduli of supercooled hard
spheres~\cite{MasonPRLxxx}. Care must be taken in applying their
results to our hard sphere samples, due to the unavoidable
uncertainties in the determination of the absolute volume
fraction~\cite{ElMasriJSTAT2009,SegreReplyXXX}. A generally accepted
approach consists in considering the separation, $\epsilon =
(\phi_c-\phi)/\phi_c$, with respect to the location $\phi_c$ of the
(apparent) divergence of the relaxation time as obtained from a mode
coupling theory fit of $\tau_{\alpha}(\phi)$ [In our work, $\phi_c
\approx 0.59$~\cite{BrambillaPRL2009}, while $\phi_c \approx 0.575$
in Ref.~\cite{MasonPRLxxx}]. For our most diluted sample, $\phi =
0.5468$ and $\epsilon \approx 0.07$, corresponding roughly to the
data for $\phi = 0.53$ in Ref.~\cite{MasonPRLxxx}. At that volume
fraction, $G'$ and $G''$ have approximately equal magnitude in all
the accessible frequency range, similarly to the case of our soft
spheres at $\phi = 0.57$. However, no spatial correlation of the
dynamics is observed in the hard spheres at $\phi = 0.5468$, while
$\xi ~\sim 2$ mm for the soft spheres at $\phi = 0.57$. This
difference is even more striking if we consider our most
concentrated hard sphere sample, for which $\xi$ is limited to a few
particle sizes~\cite{BrambillaPRL2009}, although the sample is
solid-like ($G'$ dominates over $G''$) and the magnitude of its
elastic modulus~\cite{NoteHS} is comparable to that of the Laponite
suspension, which exhibits systems-size dynamical correlations.

Our measurements thus indicate that, while elasticity is a
necessary condition for observing extended spatial correlations of
the dynamics, it is not a sufficient one. The microscopic origin of
the elasticity must also play a crucial role: the differences
discussed above suggest that $\xi$ is modest in systems where the
elasticity has an entropic origin, as for the hard spheres, and no
long-lived network of contacts exists. By contrast, $\xi$ is very
large in systems where elasticity has an enthalpic origin, be it due
to the bending rigidity of the protein filament network in
``artificial skin'' and the backbone in diluted gels made of
strongly attractive particles, or to the bulk elasticity of squeezed
particles, as in the soft spheres and the onions. Note that the
Laponite suspension falls also in the latter category: while
the platelets are not in direct contact, they interact via a screened
Coulombic repulsive potential and can thus be effectively regarded
as squeezed soft particles. Interestingly, recent
work~\cite{DelgadoPRL2009} has identified a structural length scale
that controls the mechanical properties of attractive colloidal systems: this
length scale is of the order of the particle size for hard spheres,
while it grows as short-ranged attractive interactions become
increasingly important, as in colloidal gels. Our data suggest that
this behavior may be mirrored by a similar growth of $\xi$. In
repulsive, squeezed systems, the same role is presumably played by a
persistent network of interparticle forces, such as that
visualized by confocal microscopy in compressed
emulsions~\cite{BrujicFaradayDiscussion2003}.

In view of the above discussion, it is natural to compare our
results to numerical and experimental work on driven jammed systems.
In these works, the sample is sheared by imposing either a
continuous deformation or a sinusoidal one. The dynamics are
quantified by the particle displacements perpendicular to the
shear direction or after subtracting the affine component (for a
continuous shear), or by comparing successive configurations at zero
deformation (for oscillatory shear). For a 2D granular
medium sheared  at a finite shear rate, Lechenault \textit{et al.}~\cite{LechenaultEPL2008} find
that a spatial correlation function analogous to our $G_4$ decays on
a length scale $\xi_4$ on the order of ten particle sizes at the
jamming transition, and that this length decreases upon further
compression, although only a restrained range of densities above
jamming could be probed, due to the particle stiffness. Simulations
of sheared soft disks~\cite{OlssonPRL2007} indicate that above the
jamming transition and at finite shear stress, $\sigma$, $\xi$
remains moderate, but a scaling analysis suggests that $\xi$
diverges in the limit $\sigma \rightarrow 0$ for all densities above
jamming. This is confirmed by quasi-static shear simulations of
jammed soft
particles~\cite{HeussingerPRL2009,Heussingercondmat2010}, where it
is shown that spatial correlations of the dynamics above jamming are
limited only by the system size.

Therefore, the dynamical behavior reported here appears to be due to
the connected nature of the materials investigated,
where a strain field can propagate over very large distances, as in
fully jammed systems sheared at a vanishingly small rate. A related issue, still open, concerns
the microscopic origin of the dynamics. The systems studied in
Refs.~\cite{LechenaultEPL2008,OlssonPRL2007,HeussingerPRL2009,Heussingercondmat2010}
are athermal, with no dynamics in the absence of an
external driving. Although the systems presented here are thermal and no external drive is applied to them, it is unlikely that
thermal motion alone is responsible for their dynamics in the jammed
state. This is illustrated, e.g., by the volume fraction dependence
of the dynamics reported for the soft spheres of
Ref.~\cite{SessomsPhilTransA2009}. In the supercooled regime, the
relaxation time, $\tau_c$, grows sharply with $\varphi$, as in hard
sphere systems. Above random close packing, a different regime sets
in, where $\tau_c$ grows very slowly with $\varphi$. Thus, above jamming,
the relaxation time is orders of magnitude smaller than what
expected by extrapolating the behavior in the supercooled regime,
strongly suggesting that an additional relaxation mechanism has set
in. We propose that, quite generally, the relaxation of
internal stress may be such a mechanism in jammed soft matter. This
is an appealing explanation, since it would be consistent with the
observed ultra-long range correlations of the dynamics. However,
this conjecture still awaits for a direct experimental proof.

\section{Conclusions}
We have presented direct measurements of the correlation length of
the slow dynamics of a variety of glassy and jammed soft systems,
obtained using  the recently introduced PCI method. This technique
allows one to measure coarse grained maps of the dynamical activity
of a sample. Its main advantage consists in the possibility of
probing motion on a very small length scale, yet for a very large
field of view. This is an attractive feature for jammed systems,
where motion is very restrained but highly correlated spatially.
Additionally, the technique does not require individual particles to
be imaged; accordingly, it can be applied to systems for which
direct visualization by optical or confocal microscopy is not
possible, including turbid samples as, e.g.,
foams~\cite{SessomsSoftMatter}.

We find that in deeply jammed systems $\xi$ is quite generally very
large, typically on the order of the system size. This is in
striking contrast with the modest correlation lengths measured in
glass formers, including colloidal hard spheres. An analysis of the
viscoelastic properties of the various systems investigated here
shows that a well developed elasticity ($G' > G''$) always
accompanies the presence of long-ranged spatial correlations of the
dynamics, regardless of the absolute magnitude of the elastic
modulus. The reverse, however, is not true, as exemplified by
concentrated hard sphere suspensions, whose macroscopic rheological
response is predominantly solid-like, whereas the correlation length
of the dynamics is limited to a few particle sizes. These results
highlight the crucial role of the microscopic origin of elasticity
(entropic vs. enthalpic) in determining the range of spatial
correlations of the dynamics. Further work will be needed to test
thoroughly these ideas, for example by varying continuously the
strength of attractions in dense colloidal suspensions, so as to
change progressively the nature of the
elasticity~\cite{DelgadoPRL2009} and, presumably, the correlation
length $\xi$.

\section{Acknowledgements}

S.M. has been supported by Unilever, G.B. by the R\'{e}gion
Languedoc Roussillon and the CNES. L.C. acknowledges the support of
the Institut Universitaire de France and of the ANR grant
``Dynhet''. The collaboration between L.C. and V.T was supported in
part by the CNRS (PICS N. 2410). V.T. and D.S. acknowledge financial
support from the Swiss National Science Foundation. We thanks L.
Berthier, S. Ciliberto and L. Ramos for numerous and illuminating discussions.


\begin{references}
\bibitem{NoteLaponite} The data $G' = 300$ Pa and $G'/G'' = 20$ taken from Ref.~\cite{XXXLaponiteElasticity} are for a suspension at 3.2 wt\%, instead of 3.5 wt\% as in our experiments. From Ref.~\cite{XXXLaponiteElasticity}, we estimate that the elastic modulus increases by a factor $\sim 1.5$ when the concentration increases from 3.2 wt\% to 3.5 wt\%.
\end{references}
\end{document}